# Differences at low *l* in Planck's first light sky map of the cosmic microwave background from WMAP's and COBE's

Short title: Differences at low *l* in Planck's first light sky map from WMAP's and COBE's

## Keith S Cover

Corresponding author:
Keith S Cover, PhD
Department of Physics and Medical Technology
VU University Medical Center
Postbus 7057
1007 MB Amsterdam
The Netherlands
Email: Keith@kscover.ca

PACS **98.80.Es** – Observational cosmology

PACS **02.30.Zz** – Inverse problems

## Abstract
The recent release of the first light sky map of the cosmic microwave background (CMB) from the Planck satellite provides an initial opportunity for comparison with the WMAP and COBE sky maps and their reconstruction algorithms. The precision of the match between Planck's and WMAP's anisotropies below several degrees in size, which corresponds to spherical harmonics with high $l$, provides confidence that the differences between the anisotropies at low $l$ are substantial. If the Planck first light sky map is taken as the gold standard, the results seem to suggest the low $l$ components of the WMAP map and a considerable part of the COBE sky map have a similar reconstruction artefact. As the Planck first light sky map covers only about 10% of the sky, any conclusions drawn from this comparison are speculative but deserving of further investigation.

**Introduction.** – Imaging of the brightness of the whole sky between about 40 GHz and 400GHz is believed to provide a "baby picture" of the universe. Referred to as the cosmic microwave background (CMB), variations in the intensity of the CMB with direction in the sky are believed to be due to the structure of the universe when it was about 400,000 years old. As these variations, which are referred to as anisotropies, are only about 1 part in 100,000 of the intensity of the CMB, they require extremely stable and reliable instrumentation and processing to detect them with high confidence.

The release of the Planck first light (PFL) sky map by the Planck team on Sept 17, 2009 was in the form of a single image covering about 10% of the sky. While the release of the image seemed to be intended primarily for publicity reasons, the high quality of the image allowed for a cautious comparison with the WMAP and COBE sky maps. While the acquisition frequency of the sky map was not provided in the release, nevertheless the image's minimal foreground suggests it was acquired at about 143GHz. As the PFL sky map only covers a fraction of the sky and is only a single image any conclusions drawn from the sky map must be considered speculative. However, as the full formal release of Planck's first year data is not due for another year or two, cautious comparison seems worthwhile.

The first satellite missions to report the detections of the anisotropies were COBE [SmootGF1992, BennettCL1996] and WMAP [BennettCL2003, HinshawG2003, HinshawG2009]. Neither satellite measured the sky maps directly. Instead, to generate sky maps of the whole sky, each satellite measured the difference between two points in the sky many times per second while each satellite was scanned over a large region of the sky. These measurements were made at multiple frequencies and polarisations simultaneously. To generate a map of the whole sky, observations from at least 1 year of measurements where reconstructed into a sky map. A sky map was generated for each frequency/polarisation pair for each satellite.

A great deal of effort went into the design of both the COBE and WMAP satellites to insure the reliability of any detection of anisotropies. The measurements were designed such that for an anisotropy to be considered detected it must show up in all of the frequency bands and in both polarisations. Launched about a decade after COBE, WMAP had more sensitive instrumentation which allowed it to acquire sky maps with 3,145,728 pixels over the whole sky as compared to COBE's 6,144. In the series of papers reporting the initial WMAP results, the smoothed version of the WMAP sky maps, which corresponded to spherical harmonics at low $l$, were shown to correspond well with the COBE sky maps [Hinshaw2003].

The COBE and WMAP image reconstruction algorithms were designed and implemented by very competent and well funded researchers [SmootG1992, BennettCL1996, BennettCL2003, HinshawG2003, HinshawG2009]. Thus any major problems with the image reconstruction of the sky maps are likely to primarily reflect problems with the current state of the art in the design and assessment of image reconstruction algorithms.

A common feature of the Planck, WMAP and COBE satellites was that all three made simultaneous measurements at multiple frequencies and polarisations. However, the design of the Planck satellite is different from WMAP's and COBE's in several ways.

First, Planck measures the sky at a single point as opposed to WMAP's and COBE's differential measurements between two points. Second, Planck repeats the measurement of a ring on the sky every minute for 1 hour. It then moves onto the next ring. Planck's trivially simple image reconstruction for each ring is to average all the revolutions together into a single ring. Thus, each point in the sky will be measured 60 times during one hour and the 60 measurements will be averaged together. Therefore, within 1 hour, Planck has a reliable measurement of the intensity of the CMB over a single ring. It takes Planck 6 months to scan the whole sky.

Several papers have raised questions about the reliability of the WMAP sky maps at low $l$ based on unusual properties of the sky maps. The improbable alignment of the quadrapole and octapoles of the sky maps, particularly with the earth's orbit around the sun, has been discussed extensively in the literature [Oliveria-CostaA2004, SchwarzDJ2004, LandK2007, LiuH2009]. These alignments are commonly referred to by the colourful term the "Axis of Evil". Another paper noted a puzzling correlation between the large-scale non-Gaussian patterns in the CMB and WMAP's observation numbers [LiTP2009].

Cover [CoverKS2009] also found a perplexing property of the image reconstruction used for the official WMAP sky maps. It was found that for each of the sky maps from WMAP's 20 channels, sky maps with no anisotropies were a better fit to the uncalibrated time ordered data (TOD) than WMAP's official sky maps. In this calculation WMAP's calibration parameters for each channel were allowed to vary. This result raised the possibility that there was something amiss with the calibration of the WMAP measurements that had a substantial impact on the official sky maps.

**Methods.** – Three sky maps were used for the analysis presented in this paper. The first was the sky map at 94 GHz from the WMAP 5 year analysis smoothed to 20 arc minutes. The second image was the same as the first, but with the PFL image overlaid in the regions of the WMAP sky map where the PFL data was available. The PFL data had also been smoothed to 20 arc minutes. The third image was the COBE sky map. It was the result of a combination of the sky maps from all of COBE's polarisations and frequency bands over COBE's full 4 year of observations.

All three of the images used in this analysis were obtained from the slide show presented by Gary Hinshaw of the WMAP team at the Bielefeld International workshop on Cosmic Structure and Evolution (Sept 23-25, 2009, Bielefeld, Germany). As part of this analysis, the validity of the comparison was double checked to confirm it was a valid comparison between the PFL image and the WMAP's sky maps. The WMAP team had overlaid the PFL image on the WMAP sky map at 94GHz. Visual comparison of the WMAP only and WMAP/Planck greyscale images indicated that anisotropies composed of spherical harmonics with high $l$ were in very good agreement between WMAP and the PFL. The slide show is available online at http://www.physik.uni-bielefeld.de/igs/cosmology2009/cosmic-ws09.html.

The first step of this analysis was to convert each of the 3 images from false colour to greyscale. Greyscale images are routinely used in medical imaging as colour sometimes de-emphasises important structures in images. The next step was to subtract the WMAP/Planck sky map from the WMAP only sky map. Figure 1 shows the WMAP only, WMAP/Planck and the difference image sky maps. The shape of the PFL sky map, which is limited to about 10% of the sky, determines the shape of the difference sky map.

A scatter plot of the Planck pixel values versus the difference pixel values can provide valuable insight into the quality of the difference image. A scatter plot of the values of the pixels for the PFL sky map versus the difference sky maps is shown in Fig 2. The log of the pixel count is plotted for each Planck/difference pixel pair.

The difference sky map was compared to the COBE sky map. The greyscale version of the COBE sky map is shown in Fig. 3(a). Before overlaying the difference sky map on the COBE sky map it received two steps of processing. First it was smoothed with a Gaussian blurring filter with a radius of 10 WMAP pixels. Second, it was multiplied by a gain and had a baseline added. The gain and baseline were chosen based on visual comparison with the COBE sky map. The COBE sky map with the difference sky map overlaid is shown in Fig 3(b).

**Results. –** Visual comparison of the WMAP (Fig 1a) and Planck sky map (Fig 1b) shows little obvious difference. However, careful inspection, while showing agreement at high $l$, suggests a difference at low $l$. The difference image (Fig 1c) shows primarily smoothly varying structures. This implies that for high $l$, the WMAP and Planck sky maps are closely matched and well calibrated. However, there are substantial differences at low $l$ between the WMAP and Planck sky maps.

The scatter plot in Fig 2. shows the amplitude of the difference sky map is about half of the anisotropies in the Planck sky map. Also, the scatter plot shows little correlation between the Planck and difference sky maps.

More careful examination of Fig. 1(c) shows the difference sky map to be primarily bright in the northern hemisphere and dark in the two regions of the southern hemisphere. Examination of Fig. 3(b) shows these bright and dark regions generally align with the corresponding bright and dark regions on the COBE sky map. More careful examination of each of these three regions shows a rough correlation of the variation of intensity within each of the regions with the corresponding locations on the COBE sky maps.

**Discussion. –** The availability of the PFL sky map provides an initial opportunity to assess the performance of the image reconstruction used to calculate the WMAP and COBE sky maps. However, it must always be kept in mind that the PFL sky map was released primarily for publicity reasons. As a consequence, the quality of the sky map is substantially smoothed from that available to the Planck team and virtually no supporting documentation is available for the image other than that generally available for Planck.

The most important characteristic of the difference image (Fig 1c) is what it does not have. The WMAP sky map clearly shows a texture on a scale of about 1 degree that corresponds to the peak of the WMAP power spectrum. The amplitude of this texture on the difference image is much smaller, indicating the PFL image has a very similar texture and calibration as the WMAP sky map. The difference image does have a fine grain speckle that is close to the pixel size of the image. This is likely do to a slight registration problems between the WMAP and PFL images.

The scatter graph shown in Fig. 2 gives an indication of the quality of the PFL sky map. It shows no correlation between the PFL values and the pixel values of the difference image. Thus, it is unlikely the difference image is due to an artefact in the PFL. This result supports the working hypothesis that the PFL sky map is of sufficient quality for some initial analysis.

If the scatter plot showed any correlation between the PFL and difference images there would be cause for concern that the difference image is some artefact of the PFL. For example, if the values of the pixels in the difference image increased with the pixel values in the PFL this would suggest the difference image contained some information from the PFL image. This additional information could have been due to, for example, reversing the colour coding of the published PFL. Since there is no correlation, this suggests the information in the difference image had little to do with the PFL image. While the scatter plot does not completely rule out the possibility that the difference image is an artefact in the PFL, it provides reassurance that this is not the case.

The initial impression that the WMAP sky map minus the PFL sky map seems to be roughly equal to the COBE sky maps suggests there is something wrong at low $l$ in one or more of the sky maps. When combined with the fact that the WMAP and COBE sky maps match up at low $l$ [HinshawG2003], the results seem to be at odds with the widely held belief that both WMAP and COBE at low $l$ are reliable sky maps of the CMB.

Planck's relatively simple scan pattern and image reconstruction makes calibration of the Planck data simpler than WMAP's or COBE's. Because of the averaging over each 1 hour ring, the calibration over the averaged ring can be characterized by two parameters – a gain and baseline. However, precise calibration between rings can be more complicated. Nevertheless, Planck's simple scan pattern should measure both high and low $l$ spherical harmonics with equal reliability.

In contrast, each year of TOD for each of WMAP's 20 channels requires 2 parameters for every hour of TOD. Thus, a total of 17,532 parameters are required per channel-year, the minimum number of WMAP calibration parameters required to generate a sky map. Therefore, as mentioned above, reliable calculation of the WMAP calibration parameters is the more challenging problem. The WMAP calibration issue may yield sky maps that have differing sensitivities to various spherical harmonics.

As outlined in the introduction, there have been concerns expressed about the accuracy of the WMAP sky map at low $l$. From an image interpretation point of view, the most widely discussed concerns have focused on the improbable alignment of the quadrapole and octapole of WMAP's sky maps, including with the earth's orbit around the sun [Oliveria-CostaA2004, SchwarzDJ2004, LandK2007, LiuH2009].

Examining the image reconstruction used in WMAP, Cover [CoverKS2009] found the perplexing result that no anisotropies were a better fit to the TOD than the official sky maps for each of the 20 channels when the calibration parameters were allowed to vary. The form of the analysis used was not implemented in a way to determine if the problem was concentrated at low or high *l*. But a reanalysis only constraining low *l* or high *l* harmonics to zero could provide useful insight into this issue.

One possible scenario that appears to account for the difference between Planck and WMAP-COBE at low *l* is suggested by the Axis of Evil. The alignment of the Axis of Evil with the earth's orbit around the sun suggests it could be due to a reconstruction artefact. Imperfect calibration of the WMAP TOD could allow some of the Doppler shifted CMB due to WMAP's orbit around the sun to contaminate the sky maps [CoverKS2009]. As WMAP's orbit tracks the earth's, it would explain the Axis of Evil's improbable alignments. As the error in the calibration is at most a few percentage points, any artefact might be a perturbation and thus added to the CMB true signal [HinshawG2003,CoverKS2009].

If Planck is considered to have measured the true signal and WMAP is considered the true signal plus artefact, then WMAP minus Planck would yield the artefact. This suggests the difference sky map is a good estimate of the artefact. The match of the difference image with the COBE sky map would then suggest a considerable part of the COBE sky is also a reconstruction artefact. Under this scenario, the similarity between WMAP and COBE at low *l* [HinshawG2003] is consistent with both having similar artefacts. But how is it possible that WMAP and COBE could have similar reconstruction artefacts?

The fact that there is a substantial difference at low *l* between the Planck and WMAP-COBE sky maps suggests the possibility that there is far less information in the WMAP and COBE TOD about the low *l* anisotropies than previously realised. This does not seem to be a problem at high *l* as Planck and WMAP give the same results about high *l* anisotropies. This is likely because they both had the same information about the CMB in their TOD at high *l* even though they used very different scanning patterns.

Both WMAP and COBE used a sophisticated scanning pattern and image reconstruction algorithm. The scanning patterns of each of the satellites only concentrated on one region of the sky at once and then gradually moved to the next. The measurement from the regions of the whole sky where then stitched together during the image reconstruction. This might have left ambiguity in the TOD with regards to the low *l* sky. In other words, a wide range of low *l* sky maps may have been consistent with the uncalibrated TOD. The process of selecting the calibration coefficients may have actually substantially constrained the range of low *l* sky consistent with the TOD. So a much smaller range of low *l* sky maps may have been consistent with the calibrated TOD than the uncalibrated TOD. For WMAP and COBE to yield similar but incorrect sky maps at low *l* would require them to introduce similar biases during their image reconstruction possibly via the choice of the calibration parameters.

How can we test the hypothesis that there is far less information in WMAP and COBE TOD about the low *l* sky than previously realized? Cover [CoverKS2009] proposed the criterion of asking how consistent WMAP's TOD were with sky maps containing no anisotropies while allowing WMAP's calibration parameters to vary. This was the first time the impact that WMAP's calibration parameters had on the WMAP sky maps was discussed in the literature other than by the WMAP team. After applying the proposed criterion Cover found that, for each of WMAP's 20 channels, a sky map the contained no anisotropies but did include the dipole, was a better fit to the TOD than the official WMAP sky map. This finding suggests the large low $l$ differences between WMAP and Planck may be due to a problem with the calibration of WMAP's TOD. This hypothesis could be tested by a more detailed application of Cover's reanalysis.

Cover's proposed criterion constrained the anisotropies to zero for all $l$ spherical harmonics and then allowed the calibration parameters to vary. The reanalysis could be repeated by first constraining the amplitude of the high $l$ spherical harmonics to zero but allowing the low $l$ harmonics, as well as the calibration parameters, to vary. If there is little information in the TOD about the low $l$ anisotropies, allowing the low $l$ spherical harmonics to vary will do little to improve the fit to the TOD as compared to no anisotropies. In contrast, constraining the low $l$ harmonics to zero while allowing the high $l$ harmonics to vary should substantially improve the fit.

The details of the spherical harmonics of the CMB obtained from the sky maps are believed to tell much about the early universe. It would be desirable to be able to extract more information about the spherical harmonics from PFL image. However, the limited fraction of the sky covered, combined with the degradation of the data due to its presentation as a PFL image, would pose barriers to reliable values.

As the PFL image was released primarily for publicity reasons it is important to consider other possible causes of the difference at low $l$ other than WMAP and COBE having a similar reconstruction artefact. The precise match at high $l$ strongly suggests the PFL image and the WMAP sky maps were properly scaled for comparison. The histogram in Fig. 2 suggests the difference image is independent of the PFL image, apparently ruling out the colour display of the PFL image as a source of the difference. The possibility that Planck has poor measurement abilities at low $l$ is judged extremely unlikely based on the care that when into the design of the satellite. Thus, given the limited information available about the PFL image, the best guess for the difference at low $l$ is judged to be a real difference between the true CMB and the WMAP and COBE sky maps at low $l$.

In light of the results of the analysis presented in this paper, the substantial differences between the WMAP and PFL sky maps at low $l$ need to be studied carefully for the rest of the sky. Also, the match between the COBE sky map and difference between the WMAP and PFL sky maps needs to be confirmed over the rest of the sky. As only about 10% of the sky was covered by the PFL image, and the publicity nature of the release of the image, the results of this study must be considered speculative.

**Acknowledgements.** - Thanks to Drs. Bob W. van Dijk, Francesco Sylos Labini and Dominik J Schwarz for their helpful comments. Discussions at the Cosmic Structure and Evolution Workshop at the University of Bielefeld (Sept 23-25, 2009) provided valuable background for this work. While much of the research in this project was conducted on the author's own time, additional support was received from the Netherlands' Virtual Laboratory for e-Science (VL-e) Project and the Department of Physics and Medical Technology of the VU University Medical Center.

# REFERENCES

[BennettCL2003] Bennett C. L. et al., *Astrophysical Journal Supplement*, **148** (2003) 1.
[BennettCL1996] Bennett C.L, et al., *Astrophysical Journal Letters*, **464** (1996) L1.
[CopiCJ2004] Copi C. J., Huterer D., Starkman G. D., *Phys. Rev. D.*, **70** (2004) 43515.
[CoverKS2006a] Cover K. S., *Rev. Sci. Instrum.*, **77** (2006a) 075101.
[CoverKS2008] Cover K. S., *Rev. Sci. Instrum.*, **79** (2008) 055106.
[CoverKS2009] Cover K. S., Europhys Let., **87 (**2009)  69003.
[FreemanPE2006] Freeman P. E., Genovese C. R., Miller C.J., Nichol R. C. Wasserman L., *Astrophysical Journal* **638**  (2006) 1.
[HinshawG2003] Hinshaw G. et al., *Astrophysical Journal Supplement*, **148** (2003) 63.
[HinshawG2009] Hinsahw G. et al, *Astrophysical Journal Supplement*, **180** (2009) 225.
[LandK2007] Land K, Magueijo J., Mon. Not. R. Astron. Soc., **378** (2007) 153.
[LiTP2009] Li T.P., et al., Mon. Not. R. Astron. Soc., 398 (2009) 47.
[LiuH2009] Liu H., Li T.P., astro-ph/0907.2731, (preprint) 2009.
[Oliverira-CostaA2004] de Oliveira-Costa A, Tegmark M, Zaldarriga M, Hamilton A, *Phys. Rev. D.,* **69** (2004) 063516.
[SchwarzDJ2004] Schwarz D. J., Starkman G. D., Huterer D., Copi C. J., *Phys. Rev. Let.*, **93** (2004) 221301.
[SmootGF1992] Smoot G. F. et al., *Astrophysical Journal Supplement*, **396** (1992) L13.
[TegmarkM1997] Tegmark M., *Phys. Rev. D.*, **56** (1997) 4515.
[WrightEL1996] Wright E. L., Hinshaw G., Bennett C.L. *Astrophysical Journal*, 458 (1996) L53.

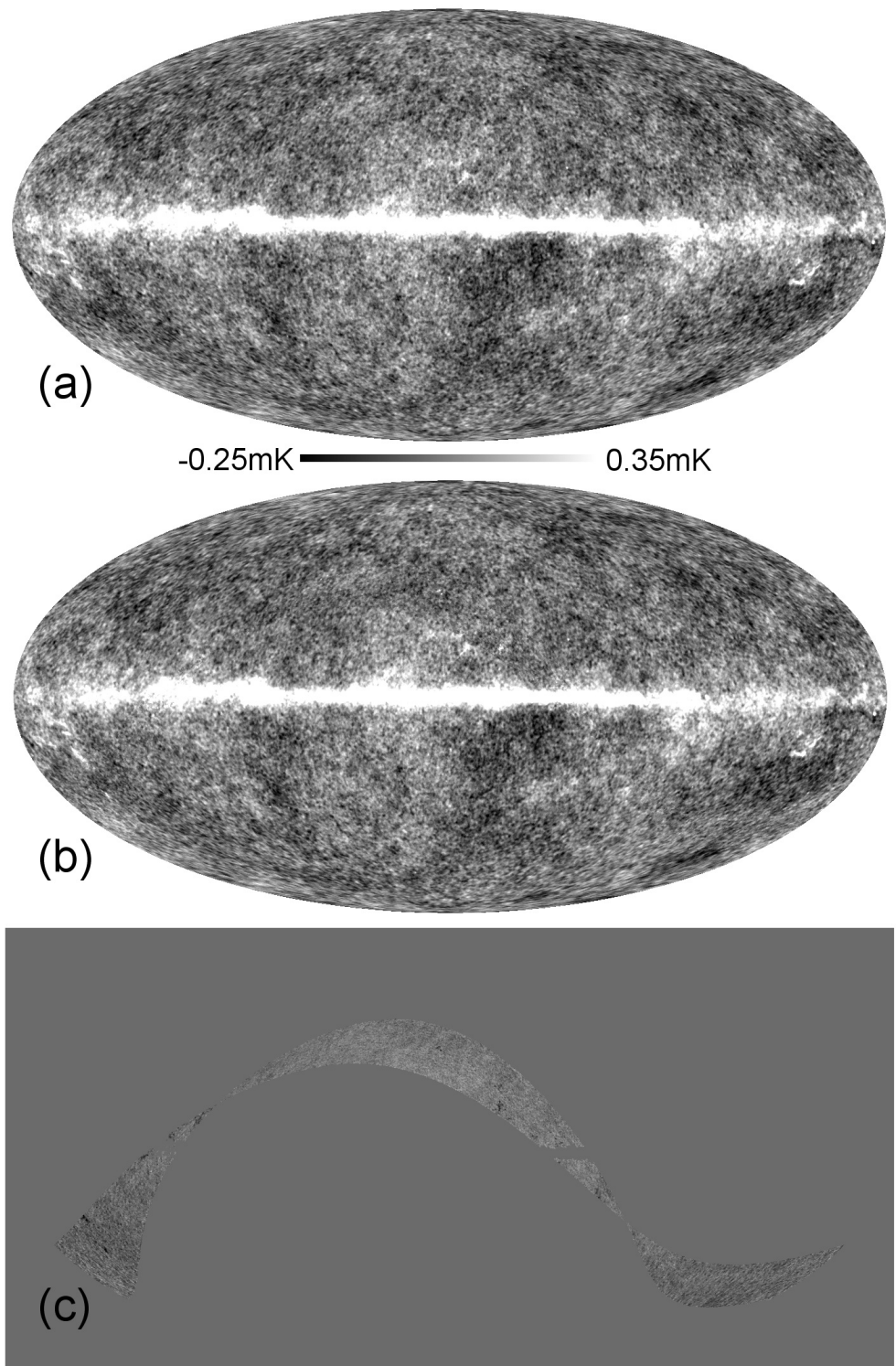

Fig. 1: (a) Official WMAP sky at 94 GHz  (b) The same sky map as (a) but with pixels replaced by those from Planck's first light sky map where available (c) difference between the WMAP and WMAP/Plank sky maps. The shape of the non-zero pixels in (c) corresponds to the region of Planck's first light image. All three images have the same brightness scale. The colour version of (a) and (b) were provided by WMAP/NASA.

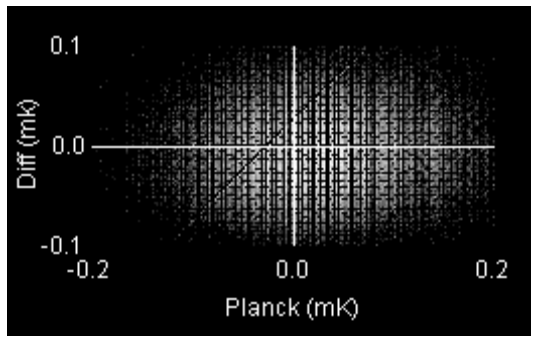

Fig 2: Scatter plot of the pixel values for the Planck first light versus the difference image.

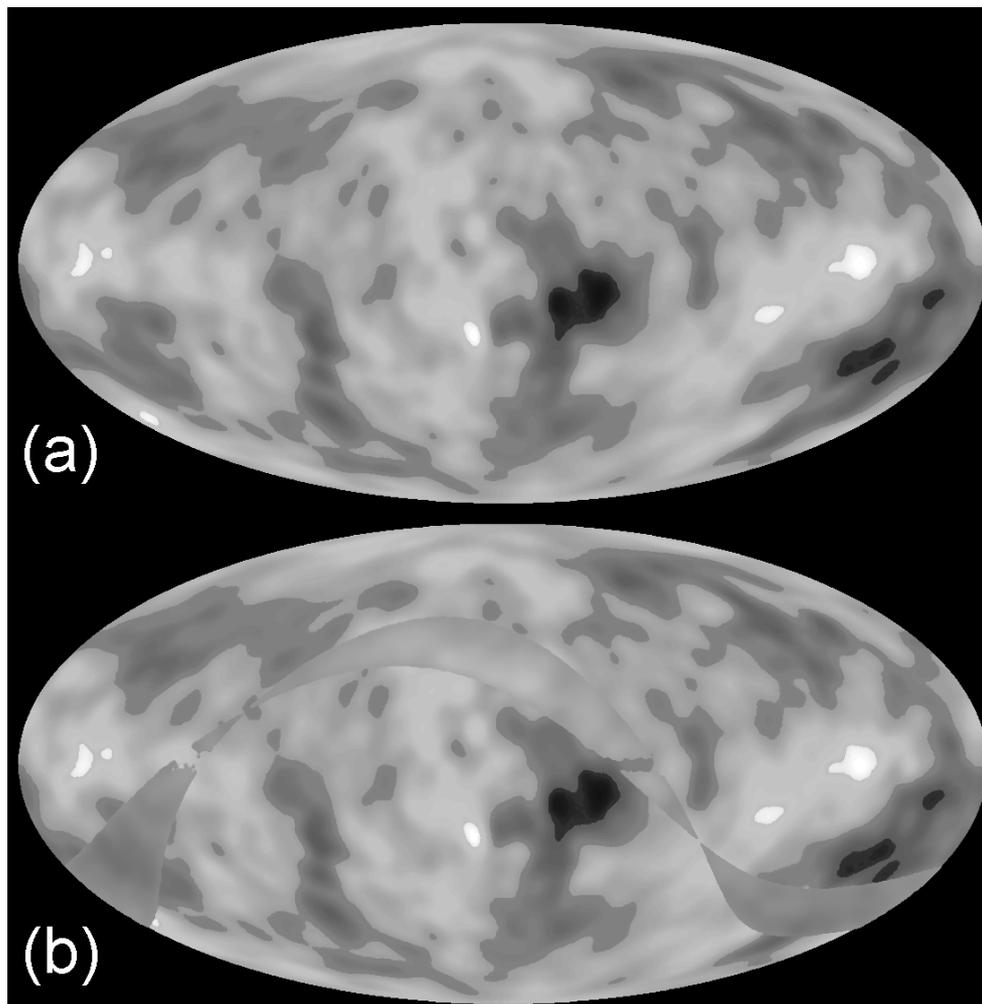

Fig. 3: (a) the COBE sky map averaged over all frequencies band.  (b) same as (a) but with a smoothed version of the difference image overlaid. The colour version of (a) was provided by WMAP/NASA.